# Matter-enhanced antineutrino flavor transformation
# and supernova nucleosynthesis

Yong-Zhong Qian[1]

*Institute for Nuclear Theory, NK-12, University of Washington, Seattle, WA 98195*

and

George M. Fuller[2]

*Physics Department, 0319, University of California, San Diego, La Jolla, CA 92093*

## Abstract

Matter-enhanced antineutrino flavor transformation between $\bar{\nu}_e$ and $\bar{\nu}_{\mu(\tau)}$ can occur in supernovae if the vacuum masses for these species satisfy $m_{\bar{\nu}_e} > m_{\bar{\nu}_{\mu(\tau)}}$. For $\delta m^2 > 1$ eV$^2$, such flavor transformation can affect the electron fraction $Y_e$ in the neutrino-heated supernova ejecta. We point out that such flavor transformation will not drive $Y_e > 0.5$ at the $r$-process nucleosynthesis epoch in the best available supernova model for such nucleosynthesis. Consequently, there is no obvious conflict between matter-enhanced antineutrino flavor transformation $\bar{\nu}_e \rightleftharpoons \bar{\nu}_{\mu(\tau)}$ and $r$-process nucleosynthesis in the neutrino-heated supernova ejecta.

---

[1] E-mail address: yzqian@phys.washington.edu.
[2] E-mail address: gfuller@ucsd.edu.

In this paper, we study matter-enhanced antineutrino flavor transformation between $\bar{\nu}_e$ and $\bar{\nu}_{\mu(\tau)}$ in supernovae. In particular, we examine the effects of such flavor transformation on the electron fraction $Y_e$ in the neutrino-heated supernova ejecta. We show that matter-enhanced flavor transformation $\bar{\nu}_e \rightleftharpoons \bar{\nu}_{\mu(\tau)}$ will not drive $Y_e > 0.5$ when $r$-process nucleosynthesis takes place in the best available supernova model for such nucleosynthesis. In contrast to the case of matter-enhanced neutrino flavor transformation $\nu_e \rightleftharpoons \nu_{\mu(\tau)}$, there is no obvious conflict between antineutrino flavor transformation $\bar{\nu}_e \rightleftharpoons \bar{\nu}_{\mu(\tau)}$ and $r$-process nucleosynthesis in the neutrino-heated supernova ejecta.

The case of matter-enhanced neutrino flavor transformation $\nu_e \rightleftharpoons \nu_{\mu(\tau)}$ and the resultant effects on $r$-process nucleosynthesis have been studied in Refs. [1] and [2]. The possible effects of matter-enhanced antineutrino flavor transformation $\bar{\nu}_e \rightleftharpoons \bar{\nu}_{\mu(\tau)}$ on $r$-process nucleosynthesis have been discussed in Ref. [3]. Reference [2] also discusses matter-enhanced antineutrino flavor transformation $\bar{\nu}_e \rightleftharpoons \bar{\nu}_{\mu(\tau)}$ in supernovae. It was pointed out in Ref. [1] that the electron fraction in the neutrino-heated supernova ejecta is determined by the characteristics of supernova neutrinos. In fact, the freeze-out value of $Y_e$ relevant for $r$-process nucleosynthesis is approximately given by the rates for the following reactions at the freeze-out radius:

$$\nu_e + n \rightleftharpoons p + e^-, \tag{1a}$$

$$\bar{\nu}_e + p \rightleftharpoons n + e^+. \tag{1b}$$

If we denote $\lambda_{\nu_e n}$ and $\lambda_{\bar{\nu}_e p}$ as the rates for the forward reactions in Eqs. (1a) and (1b), then the freeze-out value of $Y_e$ is given by

$$Y_e \approx \frac{\lambda_{\nu_e n}(r_{\text{fo}})}{\lambda_{\nu_e n}(r_{\text{fo}}) + \lambda_{\bar{\nu}_e p}(r_{\text{fo}})}. \tag{2}$$

In Eq. (2), the freeze-out radius $r_{\text{fo}}$ satisfies

$$\lambda_{\nu_e n}(r_{\text{fo}}) + \lambda_{\bar{\nu}_e p}(r_{\text{fo}}) \approx v(r_{\text{fo}})/r_{\text{fo}}, \quad (3)$$

where $v(r_{\text{fo}})$ is the outflow velocity of the ejecta at radius $r_{\text{fo}}$. The rates $\lambda_{\nu_e n}$ and $\lambda_{\bar{\nu}_e p}$ decrease with increasing radius. Above the freeze-out radius, these rates become smaller than the rate at which the ejecta are flowing out. Consequently, $\nu_e$ and $\bar{\nu}_e$ stop interacting effectively with the free nucleons in the ejecta above the freeze-out radius.

In writing down the above equations, we have neglected the reverse reactions in Eqs. (1a) and (1b). This is because the reverse reaction rates depend sensitively on the material temperature. At the freeze-out radius, the material temperature is low and these rates are small compared with the forward reaction rates.

To calculate the rates $\lambda_{\nu_e n}$ and $\lambda_{\bar{\nu}_e p}$, we need the angular and energy distributions of the supernova neutrino fluxes. Because the freeze-out radius is sufficiently large, we can make the approximation that supernova neutrinos are emitted from a neutrino sphere. This approximation gives a good description of the angular distribution of the neutrino fluxes at large radii.

The cross sections for the forward reactions in Eqs. (1a) and (1b) are given by

$$\sigma_{\nu_e n} \approx 9.6 \times 10^{-44} \left( \frac{E_{\nu_e} + \Delta_{\text{np}}}{\text{MeV}} \right)^2 \text{ cm}^2, \quad (4a)$$

and

$$\sigma_{\bar{\nu}_e p} \approx 9.6 \times 10^{-44} \left( \frac{E_{\bar{\nu}_e} - \Delta_{\text{np}}}{\text{MeV}} \right)^2 \text{ cm}^2, \quad (4b)$$

respectively, where $\Delta_{\text{np}} \approx 1.293$ MeV is the neutron-proton mass difference. The dependence of $\sigma_{\nu_e n}$ on $\nu_e$ energy $E_{\nu_e}$ in Eq. (4a) is almost exact due to the Coulomb focusing effect for the final state charged particles. The dependence of $\sigma_{\bar{\nu}_e p}$ on $\bar{\nu}_e$ energy $E_{\bar{\nu}_e}$ in Eq. (4b) is accurate when the positron in the final state is extremely relativistic. Because typical supernova neutrino energies are on the order of 10 MeV, this is almost always the case.

From Eqs. (4a) and (4b), we note that analytic fits to the supernova neutrino energy distributions are appropriate for evaluating $\lambda_{\nu_e n}$ and $\lambda_{\bar{\nu}_e p}$ if they can reproduce the first and second energy moments, $\langle E_\nu \rangle$ and $\langle E_\nu^2 \rangle$, for individual neutrino flavors. In this case, the neutrino luminosity $L_\nu$ can serve as an overall normalization for the individual neutrino flux. The values of $L_\nu$, $\langle E_\nu \rangle$, and $\langle E_\nu^2 \rangle$ for $\nu_e$, $\bar{\nu}_e$, and $\bar{\nu}_\mu$ are given in Table 1 for three different times relevant for supernova $r$-process nucleosynthesis: tpb (time post bounce) = 6, 10, and 16 s. The values of $L_\nu$, $\langle E_\nu \rangle$, and $\langle E_\nu^2 \rangle$ for $\nu_\mu$, $\nu_\tau$, and $\bar{\nu}_\tau$ are approximately the same as those for $\bar{\nu}_\mu$. These values correspond to the neutrino characteristics obtained in the best available supernova model for $r$-process nucleosynthesis [4].

With these parameters, we can fit the local differential neutrino flux at radius $r$ to the form:

$$d\phi_\nu \approx \frac{L_\nu}{4\pi r^2} \frac{1}{F_3(\eta_\nu)T_\nu^4} \frac{E_\nu^2}{\exp(E_\nu/T_\nu - \eta_\nu) + 1} dE_\nu. \tag{5}$$

In Eq. (5), $F_3(\eta_\nu)$ is the rank 3 Fermi integral of argument $\eta_\nu$. The values of the fitting parameters $T_\nu$ and $\eta_\nu$ are also given in Table 1.

From Table 1 and Eqs. (2), (4a), (4b), and (5), we can see that the freeze-out value of $Y_e$ is always less than 0.5 in the absence of neutrino flavor transformation. We give these values of $Y_e$ as $Y_e^a$ in Table 1. The condition $Y_e < 0.5$ is necessary for $r$-process nucleosynthesis to occur. We also note from Table 1 that the energy distributions for the $\nu_e$, $\bar{\nu}_e$, and $\nu_\mu$ or $\bar{\nu}_\mu$ neutrino fluxes are very different. As a result, the freeze-out value of $Y_e$ can be affected by transformation between different neutrino flavors. Since the $\nu_\tau$ and $\bar{\nu}_\tau$ fluxes have identical energy distributions to the $\nu_\mu$ and $\bar{\nu}_\mu$ fluxes, we will discuss the flavor transformation $\nu_e \rightleftharpoons \nu_\mu$ and $\bar{\nu}_e \rightleftharpoons \bar{\nu}_\mu$ as specific examples. It is to be understood that our conclusions regarding the effects of such flavor transformation on $r$-process nucleosynthesis are also valid for the flavor transformation $\nu_e \rightleftharpoons \nu_\tau$ and $\bar{\nu}_e \rightleftharpoons \bar{\nu}_\tau$.

References [1] and [2] show that matter-enhanced flavor transformation between a light $\nu_e$ and a $\nu_\mu$ with a cosmologically interesting mass ($m_{\nu_\mu} \approx 1 - 100$ eV) can significantly change the freeze-out value of $Y_e$. In fact, it is not necessary to have full conversion between $\nu_e$ and $\nu_\mu$ to drive $Y_e > 0.5$ [1,2]. If $r$-process nucleosynthesis comes from the neutrino-heated supernova ejecta, then the required condition $Y_e < 0.5$ places severe constraints on the mixing between $\nu_e$ and $\nu_\mu$.

However, the flavor transformation $\nu_e \rightleftharpoons \nu_\mu$ can be enhanced in supernovae only if $\nu_\mu$ is heavier than $\nu_e$. So far there is no experimental evidence to confirm that this is the actual mass hierarchy. In fact, the current experimental upper limit on the $\bar{\nu}_e$ mass is 7.2 eV [5]. Therefore, the possibility of $\nu_e$ and $\bar{\nu}_e$ being the desired hot dark matter in some cosmological models [6] has not been ruled out yet. If the actual mass hierarchy is $m_{\bar{\nu}_e} > m_{\bar{\nu}_\mu}$, then the flavor transformation $\bar{\nu}_e \rightleftharpoons \bar{\nu}_\mu$ can be enhanced in supernovae, whereas the flavor transformation $\nu_e \rightleftharpoons \nu_\mu$ is suppressed.

To illustrate the effects of the flavor transformation $\bar{\nu}_e \rightleftharpoons \bar{\nu}_\mu$, we plot the differential $\bar{\nu}_e$ capture rate on protons $d\lambda/dE_\nu = \sigma_{\bar{\nu}_e p} d\phi_\nu/dE_\nu$ with respect to $\bar{\nu}_e$ energy in Fig. 1. Since the differential capture rate is a radius-dependent quantity, the scale for the ordinate is arbitrary. The solid line in Fig. 1 corresponds to the original $\bar{\nu}_e$ energy distribution at tpb = 6 s, while the dashed line represents the case where the $\bar{\nu}_e$ assume the same energy distribution as the $\bar{\nu}_\mu$ at tpb = 6 s.

From Fig. 1, we can see that the differential capture rate peaks at a neutrino energy of $E_\nu \approx \langle E_\nu^2 \rangle / \langle E_\nu \rangle$. This energy where the rate peaks is unique for each energy distribution. Between the peaks of the differential capture rates for the two energy distributions, the solid line and the dashed line cross at a neutrino energy of $E_\nu \approx 25$ MeV. Although Fig. 1 is constructed for tpb = 6 s, we find that the generic features, especially the crossing of the differential capture rates for the two energy distributions at $E_\nu \approx 25$ MeV, as shown in Fig. 1 are also representative of the energy distributions throughout the period from tpb = 6 s to 16 s of the $r$-process nucleosynthesis epoch.

It is obvious from Fig. 1 that the worst effect of the flavor transformation $\bar{\nu}_e \rightleftharpoons \bar{\nu}_\mu$ on $Y_e$ occurs when antineutrinos with energies less than the crossing energy $E_\nu \approx 25$ MeV are fully transformed, whereas those with energies greater than the crossing energy remain unchanged. This is because there are more $\bar{\nu}_e$ than $\bar{\nu}_\mu$ with energies $E_\nu < 25$ MeV. This worst case obtains, for example, when antineutrinos with energies $E_\nu < 25$ MeV go through adiabatic flavor transformation and those with energies $E_\nu > 25$ MeV do not have resonances below the freeze-out radius of $Y_e$. This worst scenario may be realized because higher energy antineutrinos go through resonances at lower densities corresponding to larger radii.

The resonance condition [7] for matter-enhanced antineutrino flavor transformation

$\bar{\nu}_e \rightleftharpoons \bar{\nu}_\mu$ is given by

$$\frac{\delta m^2}{2E_\nu} \cos 2\theta = \sqrt{2} G_F(n_e + n_\nu^{\text{eff}}), \qquad (6a)$$

where $\delta m^2$ is the vacuum mass-squared difference, and $\theta$ is the vacuum mixing angle. We choose $\delta m^2 > 0$ so that $\delta m^2 \approx m_{\bar{\nu}_e}^2 - m_{\bar{\nu}_\mu}^2$ for $\theta \ll 1$. The net electron number density $n_e$ in Eq. (6a) is given by

$$n_e \equiv n_{e^-} - n_{e^+} = Y_e \rho N_A, \qquad (6b)$$

where $\rho$ is the matter density and $N_A$ is the Avogadro's number.

The effective neutrino number density $n_\nu^{\text{eff}}$ in Eq. (6a) represents the neutrino-neutrino forward scattering contributions to the neutrino propagation Hamiltonian [2]. It is defined to be

$$n_\nu^{\text{eff}} \equiv n_{\nu_e}^{\text{eff}}(r) - n_{\bar{\nu}_e}^{\text{eff}}(r) - \left[ n_{\nu_\mu}^{\text{eff}}(r) - n_{\bar{\nu}_\mu}^{\text{eff}}(r) \right], \qquad (6c)$$

where, for example, $n_{\nu_e}^{\text{eff}}(r)$ is the effective $\nu_e$ number density at radius $r$.

In the case of matter-enhanced antineutrino flavor transformation, the flavor contents of the $\nu_e$ and $\nu_\mu$ fluxes are essentially unaffected for $\theta \ll 1$. With the approximation that neutrinos are emitted from a neutrino sphere at radius $R_\nu$, the effective $\nu_e$ number density $n_{\nu_e}^{\text{eff}}$ for a radially propagating neutrino is given by

$$n_{\nu_e}^{\text{eff}} \approx \frac{L_{\nu_e}}{\langle E_{\nu_e} \rangle} \frac{1}{4\pi r^2 c} \frac{R_\nu^2}{4r^2}, \qquad (6d)$$

where $c$ is the speed of light. The effective $\nu_\mu$ number density $n_{\nu_\mu}^{\text{eff}}$ has an expression similar to Eq. (6d). The neutrino sphere radius at the $r$-process nucleosynthesis epoch is $R_\nu \approx 10$ km. However, the effective $\bar{\nu}_e$ and $\bar{\nu}_\mu$ number densities $n_{\bar{\nu}_e}^{\text{eff}}$ and $n_{\bar{\nu}_\mu}^{\text{eff}}$ are affected by the matter-enhanced antineutrino flavor transformation. They have to be determined from the flavor evolution history of individual antineutrinos with different energies.

In the worst scenario given above, the freeze-out value of $Y_e$ will be larger than the original value $Y_e^a$. However, for the neutrino energy distributions obtained in the best available model for $r$-process nucleosynthesis [4], we find that the freeze-out values of $Y_e$ in the worst scenario can never exceed 0.5 during the period from tpb = 6 s to 16 s when $r$-process nucleosynthesis takes place. We give these values of $Y_e$ as $Y_e^b$ in Table 1.

Of course, as discussed earlier, the worst scenario can happen only if antineutrinos with energies $E_\nu > 25$ MeV do not go through resonances below the radius where the value of $Y_e$ freezes out. More specifically, the realization of this scenario depends on the matter density and effective neutrino number density at the freeze-out radius of $Y_e$ for a particular $\delta m^2$ [cf. Eq. (6a)].

The freeze-out radius is approximately determined by Eq. (3). Since even in the worst scenario the total $\bar{\nu}_e$ capture rate $\lambda_{\bar{\nu}_e p}$ is still greater than $\lambda_{\nu_e n}$ [cf. Eq. (2) and Table 1], we may conservatively approximate $\lambda_{\nu_e n} + \lambda_{\bar{\nu}_e p} \approx 2\lambda_{\nu_e n}$. Together with Eqs. (4a) and (5), this gives

$$r_{\mathrm{fo}} v(r_{\mathrm{fo}}) \approx 10^{14} \mathrm{\ cm^2\ s^{-1}} \qquad (7a)$$

for the $r$-process nucleosynthesis epoch.

One virtue of the best supernova model adopted for the $r$-process nucleosynthesis of Ref. [4] and in this paper is that it predicts the right amount of $r$-process material ejected from each supernova. In the context of galactic chemical evolution calculations the adopted supernova model accounts nicely for the solar abundances of $r$-process elements [4]. The mass ejection rate $\dot{M}$ at radius $r$ is given by

$$\dot{M} = 4\pi r^2 \rho v. \tag{7b}$$

This mass ejection rate stays at about $10^{-6} M_\odot$ s$^{-1}$ from tpb = 6 s to 16 s in the adopted supernova model [4].

We can rewrite Eqs. (7a) and (7b) as

$$r_{\rm fo} \rho(r_{\rm fo}) \approx 2 \times 10^{12} \text{ g cm}^{-2}. \tag{7c}$$

From Eq. (7c), we can see that $\rho(r_{\rm fo}) < 2 \times 10^6$ g cm$^{-3}$ because the freeze-out radius always lies above the neutrino sphere, i.e., $r_{\rm fo} > R_\nu \approx 10$ km. In the actual supernova model, the freeze-out radius stays above 30 km. So the matter density at the freeze-out radius is below $7 \times 10^5$ g cm$^{-3}$. If we neglect the effective neutrino number density $n_\nu^{\rm eff}$ in the resonance condition given by Eq. (6a), we find that antineutrinos with energies $E_\nu > 25$ MeV will not go through resonances below the freeze-out radius for $\delta m^2 < 1$ eV$^2$.

The effective neutrino number density is approximately $20 - 30\%$ of the net electron number density at the freeze-out radius. As is evident from Eq. (6a), the effective neutrino number density tends to push the resonance position towards a larger radius for a given value of $\delta m^2$. This is to be contrasted with the opposite effects of the effective neutrino number density on matter-enhanced *neutrino* flavor transformation. As discussed in Ref. [2], for the case of matter-enhanced neutrino flavor transformation, the effective neutrino number density tends to draw the resonance position towards a smaller radius for a given $\delta m^2$. Taking into account the effective neutrino number density, we find that antineutrinos with energies $E_\nu < 25$ MeV will always have a resonance below the freeze-out radius if $\delta m^2 > 2$ eV$^2$.

The scenario which gives the worst possible effect on $Y_e$ can then be realized for adiabatic antineutrino flavor transformation with $\delta m^2 \approx 2$ eV$^2$. For any other flavor evolution scenarios, the freeze-out value of $Y_e$ will always be smaller than $Y_e^b$ in the worst scenario. It is interesting to observe that for a slightly larger $\delta m^2 \approx 3$ eV$^2$, antineutrinos with energies $E_\nu < 35$ MeV will go through resonances below the freeze-out radius. If we further assume that these antineutrinos are fully transformed, then the freeze-out value of $Y_e$ is smaller than or very close to the original value $Y_e^a$. We give these values of $Y_e$ as $Y_e^c$ in Table 1. For $\delta m^2 > 3$ eV$^2$, matter-enhanced antineutrino flavor transformation can decrease $Y_e$ below $Y_e^a$. This is because at energies $E_\nu > 25$ MeV, more $\bar{\nu}_\mu$ are transformed into $\bar{\nu}_e$ than $\bar{\nu}_e$ are transformed into $\bar{\nu}_\mu$. This effect can be seen from Fig. 1.

In the above discussion, we only have considered the possible effects of matter-enhanced antineutrino flavor transformation $\bar{\nu}_e \rightleftharpoons \bar{\nu}_\mu$ on $r$-process nucleosynthesis in supernovae. However, if the mass hierarchy $m_{\bar{\nu}_e} > m_{\bar{\nu}_\mu}$ obtains, there will be other consequences of the flavor transformation $\bar{\nu}_e \rightleftharpoons \bar{\nu}_\mu$. In particular, the flavor transformation $\bar{\nu}_e \rightleftharpoons \bar{\nu}_\mu$ taking place in a galactic supernova could cause observable effects in future neutrino detectors such as super Kamiokande. This is because the flavor transformation $\bar{\nu}_e \rightleftharpoons \bar{\nu}_\mu$ can affect the energy distribution of $\bar{\nu}_e$ detected on earth.

In fact, the predecessor of super Kamiokande, the Kamiokande II detector, detected 11 neutrino events from SN1987A, while 8 events were detected by the now inactive IMB detector. There have been many studies in the literature which try to extract information about supernova neutrinos from these 19 (mainly $\bar{\nu}_e$) events. Before we give a brief discussion of these previous studies, we emphasize that the period of the supernova process relevant for supernova neutrino detection is much earlier than the $r$-process nucleosynthesis epoch. The majority of the neutrino events in a detector come from the first a few seconds after core bounce (i.e., tpb $< 2$ s), whereas the $r$-process nucleosynthesis takes place at tpb $> 6$ s. Both the neutrino characteristics (e.g., luminosities and energy distributions) and the dynamic aspects of the supernova (e.g., density structure and hydrodynamic instabilities) are very different for these two periods. The discrepancy between $\langle E_{\bar{\nu}_e} \rangle$ and $\langle E_{\bar{\nu}_\mu} \rangle$ at earlier times is smaller than as shown in Table 1 for tpb $\geq 6$ s. The neutrino sphere radius shrinks significantly from $R_\nu \approx 50$ km at tpb $\sim 0.1$ s to $R_\nu \approx 10$ km at tpb $> 3$ s. Consequently, treatment of the flavor transformation $\bar{\nu}_e \rightleftharpoons \bar{\nu}_\mu$ and its implications will also be different.

A careful statistical analysis of the SN1987A neutrino signals has been carried out in Ref. [8]. However, the results presented as Figs. 5 and 6 in Ref. [8] were based on an overly-simplified model of supernova neutrino emission. To be specific, it was assumed that (1) the $\bar{\nu}_e$ energy distribution is given by a Fermi-Dirac distribution with zero chemical potential; (2) the temperature characterizing the Fermi-Dirac distribution decreases exponentially with time; (3) the neutrino sphere radius is fixed; and (4) the $\bar{\nu}_e$ luminosity is given by the blackbody radiation law corrected for the Fermi-Dirac statistics, i.e., there exists a specific relation between neutrino luminosity, neutrino temperature, and neutrino sphere radius. These assumptions differ from the approximations for supernova neutrino fluxes made in this paper. While we employ a Fermi-Dirac distribution with finite chemical potential for the *normalized* neutrino energy distribution, we treat the neutrino luminosity as an independent quantity [see Eq. (5)]. There is no simple analytic function which can adequately describe the evolution of neutrino luminosity and neutrino energy distributions simultaneously. In addtion, the above assumption (2) contradicts the physical effects of neutronization on the $\bar{\nu}_e$ opacities. Protons in the core provide an important source for the $\bar{\nu}_e$ opacity through the forward reaction in Eq. (1b). The lower opacity then hardens the $\bar{\nu}_e$ energy distribution as the core becomes more deficient in protons with time.

Given the poor statistics of the SN1987A neutrino events, and the above mentioned errors in the supernova neutrino emission model used to analyze these events, it is clear that conclusions regarding the $\bar{\nu}_e$ energy distribution so obtained must be taken with great caution. With a total of only 19 events, one can hardly expect more than a confirmation of the gross energetics and qualitative features of supernova neutrino emission. With the above mentioned caveats on the underlying supernova neutrino emission model used to analyze the SN1987A data, fits to the $\bar{\nu}_e$ temperature range between $\sim 3$ and $\sim 6.5$ MeV [8]. This range is clearly too broad to preclude the possible occurrence of flavor transformation $\bar{\nu}_e \rightleftharpoons \bar{\nu}_\mu$ in SN1987A.

It is our sincere hope that nature will grant us the opportunity to detect neutrinos from a galactic supernova when super Kamiokande is in full operation. In that case, we will have many more neutrino events, which may enable us to extract details of supernova neutrino energy distributions.

In conclusion, we have studied the effects of matter-enhanced antineutrino flavor transformation on the freeze-out value of $Y_e$ in the neutrino-heated supernova ejecta. We find that such flavor transformation can never drive $Y_e > 0.5$ when $r$-process nucleosynthesis takes place in the best available supernova model for such nucleosynthesis. For $\delta m^2 > 3$ eV$^2$, matter-enhanced antineutrino flavor transformation can even decrease $Y_e$ below the original value $Y_e^a$ for no flavor transformation. While the actual effects of different $Y_e$ from $Y_e^a$ on the $r$-process nucleosynthesis still await further detailed study, there is no obvious conflict between matter-enhanced antineutrino flavor transformation and $r$-process nucleosynthesis in the neutrino-heated supernova ejecta.

We acknowledge Jim Wilson for providing us with the details of his numerical supernova model used for $r$-process nucleosynthesis. This work was supported by the Department of Energy under Grant No. DE-FG06-90ER40561 at the Institute for Nuclear Theory and by NSF Grant No. PHY-9121623 and an IGPP minigrant at UCSD.

## Figure Caption

**Fig. 1** Differential $\bar{\nu}_e$ capture rate on proton with respect to $\bar{\nu}_e$ energy. The solid line corresponds to the original $\bar{\nu}_e$ energy distribution at tpb = 6 s. The dashed line obtains when $\bar{\nu}_e$ assume the same energy distribution as $\bar{\nu}_\mu$ at tpb = 6 s.

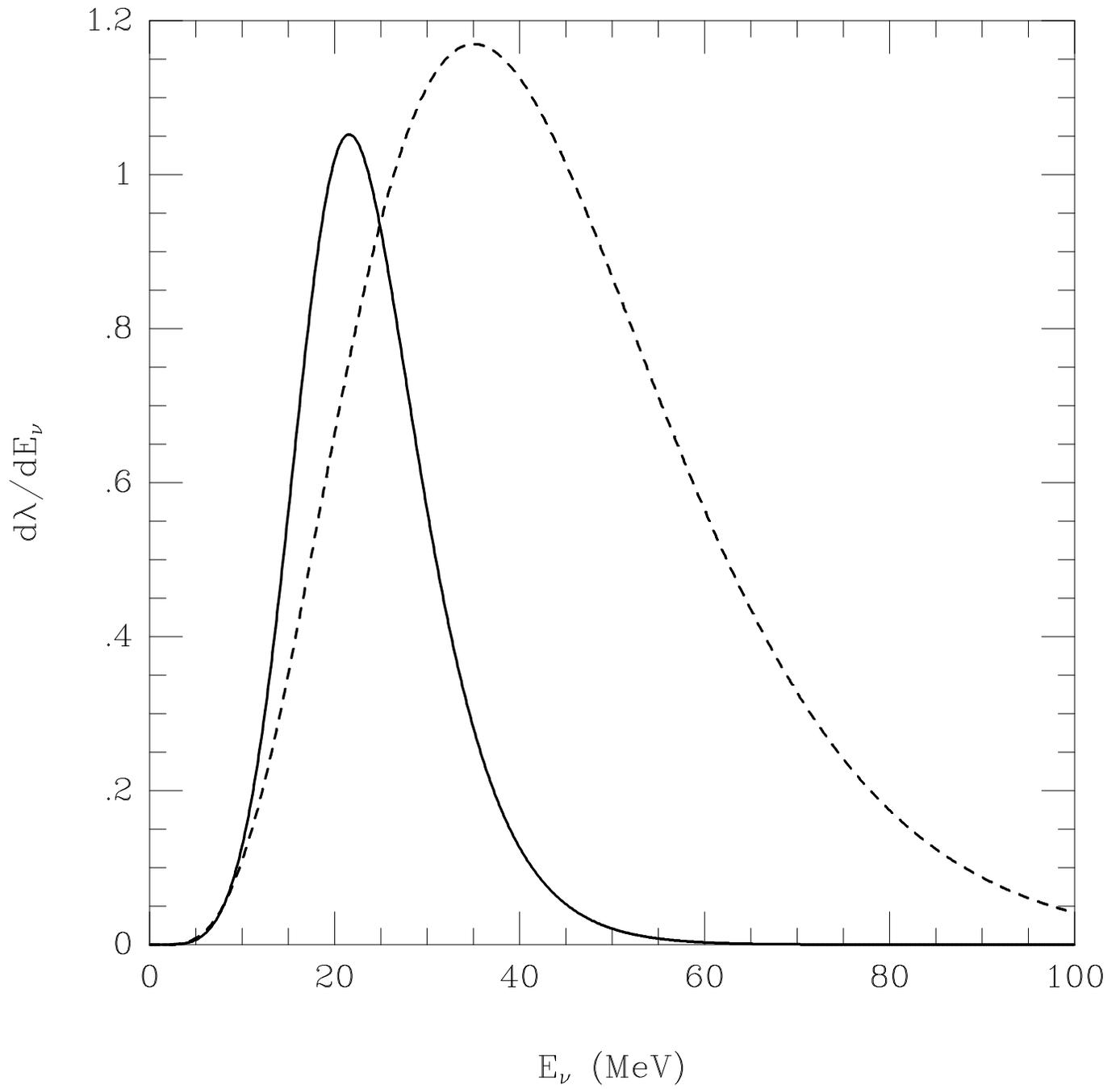

Fig. 1

Table 1: Supernova neutrino characteristics and freeze-out values of $Y_e$ for different scenarios of antineutrino flavor transformation at three representative times during the $r$-process nucleosynthesis epoch.

| tpb (s) | $\nu$ | $L_\nu$ ($10^{51}$ erg s$^{-1}$) | $\langle E_\nu \rangle$ (MeV) | $\langle E_\nu^2 \rangle$ (MeV$^2$) | $T_\nu$ (MeV) | $\eta_\nu$ | $Y_e^a$ | $Y_e^b$ | $Y_e^c$ |
|---|---|---|---|---|---|---|---|---|---|
| 6 | $\nu_e$ | 0.98 | 10.5 | 133.5 | 2.69 | 2.78 | 0.44 | 0.48 | 0.42 |
| | $\bar{\nu}_e$ | 1.05 | 18.4 | 389.6 | 3.69 | 4.97 | | | |
| | $\bar{\nu}_\mu$ | 1.64 | 26.1 | 901.1 | 8.56 | -1.27 | | | |
| 10 | $\nu_e$ | 0.57 | 10.2 | 125.1 | 2.55 | 3.03 | 0.40 | 0.41 | 0.34 |
| | $\bar{\nu}_e$ | 0.64 | 19.5 | 440.6 | 4.08 | 4.60 | | | |
| | $\bar{\nu}_\mu$ | 1.26 | 26.2 | 915.0 | 8.73 | -4.72 | | | |
| 16 | $\nu_e$ | 0.49 | 10.1 | 124.4 | 2.66 | 2.53 | 0.37 | 0.41 | 0.37 |
| | $\bar{\nu}_e$ | 0.62 | 19.6 | 449.2 | 4.31 | 4.16 | | | |
| | $\bar{\nu}_\mu$ | 0.90 | 26.6 | 937.4 | 8.75 | -1.51 | | | |